\documentclass[12pt]{iopart}
\usepackage{iopams}
\usepackage{graphicx}
\usepackage{setstack}

\usepackage{array}

\begin{document}

\newcommand{\hh}{\mathbb{H}}
\newcommand{\F}{\mathcal{F}}
\newcommand{\cc}{\mathbb{C}}

\title{Universal behavior of a bipartite fidelity at quantum criticality}
\author{J\'er\^ome Dubail$^{1}$ and Jean-Marie St\'ephan$^{2}$}

\address{$^1$ Department of Physics, Yale University, New Haven, CT 06520\\
$^2$ Institut de Physique Th\'eorique, CEA, IPhT, CNRS, URA 2306, F-91191 Gif-sur-Yvette, France
}

\eads{\mailto{jerome.dubail@yale.edu}, \mailto{jean-marie.stephan@cea.fr}}

\begin{abstract}
We introduce the (logarithmic) bipartite fidelity of a quantum system $A\cup B$ as the (logarithm of the) overlap between
 its ground-state wave function and the ground-state one would obtain if the interactions between two complementary subsystems
 $A$ and $B$ were switched off. We argue that it should
typically satisfy an area law in dimension $d>1$. In the case of one-dimensional quantum critical points (QCP)
 we find that it admits a universal scaling form $\sim \ln \ell$, where $\ell$ is the typical size of the smaller subsystem.
 The prefactor is proportional to the central charge $c$ and depends on the geometry.
We also argue that this quantity can be useful to locate quantum phase transitions, allows
 for a reliable determination of the central charge, and in general exhibits various properties that are similar
 to the entanglement entropy. Like the entanglement entropy,
it contains subleading universal terms in the case of a 2D conformal QCP. 
 
\end{abstract}

\date{\today}

\maketitle

{\it Introduction}.---
A major challenge in the study of quantum many-body systems in condensed matter physics
is the understanding and characterization of new exotic phases of matter, such as quantum critical or topological phases.
 For this purpose, various quantities and concepts have been introduced, some coming
 from quantum information theory. Amongst them, one of the most
heavily studied is the entanglement entropy (EE)\cite{eereviews}, defined through a bipartition of a total system $A\cup B$,
 usually in a pure state $|\psi\rangle$:
 \begin{equation}
 S=-{\rm Tr }\rho_A \ln \rho_A\quad,\quad \rho_A={\rm Tr}_B \,|\psi\rangle \langle \psi|
\end{equation}
The EE of a ground state is known to be universal at one-dimensional quantum critical point (QCP) \cite{hlw,vlrk,cc},
 and the leading term allows for an accurate determination of the central charge.
In higher dimension $d>1$, it obeys an area law (with possible logarithmic corrections \cite{noarealaw}):
 if $L$ is the typical size of the smaller subsystem, then $S$
scales as $L^{d-1}$.
 In this case, subleading terms\cite{kplw06,fm06,hzs07,mfs,stephan09} encode
 universal features of the system, characterizing quantum criticality or
 topological order.
 Despite all of these theoretical works, connecting the EE
 to {\it experimentally measurable} quantities remains a formidable task. One reason for that
 is that characterization of entanglement in a many-particle system generically requires the measurement of a prohibitively large number of observables. 

A different class of quantities is the one of overlaps, or fidelities. 
The idea is perhaps more intuitive and can be traced back
 to Anderson's orthogonality catastrophe \cite{Anderson}. Let $H(\lambda)$ be a
 Hamiltonian which depends on a physical parameter $\lambda$ that can be varied.
 If $\left| \lambda \right>$ is its ground state, the fidelity is
\begin{equation}
 f(\lambda,\lambda')=\left|\left\langle \lambda| \lambda'\right\rangle \right|^2.
\end{equation}
Close to a QCP the
fidelity susceptibility $\chi(\lambda) = \left( \partial_{\lambda'}^{\,2}f(\lambda,\lambda') \right)_{\lambda'=\lambda}$
 diverges, so this quantity can be used to detect quantum phase transitions \cite{fidelity,fidelity2,fidelitygu}.
The scaling behavior of the fidelity has been studied in various systems both analytically and 
numerically \cite{vsz,aletetal}. Overlaps are also interesting 
when considering time-evolutions. 
Starting from an initial state $\left| \Psi(0) \right>$, one can
 ask what the overlap of the wavefunction with the initial state after time $t$ is.
 The Loschmidt echo $\mathcal{L}(t) = | \left< \psi(0) | \psi(t) \right> |^2$ \cite{Loschmidt,Loschmidtreview} has
 been studied in connection with NMR experiments \cite{Loschmidtexp},
 and also in the context of quantum criticality \cite{Loschmidt1,Loschmidt3}.

In this letter, we introduce an overlap which shares some common properties with the EE 
(in particular we keep the idea of cutting the system into two parts),
 despite being conceptually simpler. 
We call it {\it logarithmic bipartite fidelity} (LBF), and claim that it provides valuable
 insights into quantum critical phenomena.
 We shall see in particular that the LBF obeys an area law in $d>1$,
 can be useful to locate QCPs, and has a universal scaling form at one-dimensional QCPs that involves the central charge $c$,
 much in the spirit of the EE.

{\it Bipartite fidelity}.---
Let us consider an extended quantum system $A\cup B$ described by the Hamiltonian
\begin{equation}
	\label{eq:totham}
 H=H_A+H_B+H_{A\cup B}^{(I)}
\end{equation}
where $[H_A,H_B]=0$ and $H_{A\cup B}^{(I)}$ contains all the interaction between $A$ and $B$.
 We denote by $|A\rangle$ (resp. $|B\rangle)$ the ground-state of $H_A$ (resp. $H_B$),
 by $|A\otimes B\rangle=|A\rangle \otimes |B\rangle$ the ground-state of $H_A+H_B$, and by
$|A\cup B\rangle$ the ground-state of $H$. We introduce the {\it bipartite fidelity}
$\big|\langle A\cup B|A\otimes B\rangle \big|^2 $, 
the overlap between the ground-state of the total Hamiltonian $H$, and the ground state of a
 Hamiltonian $H_A+H_B$ where all
interactions between $A$ and $B$ have been switched off.
 A more physical way of looking at this quantity is to interpret it as a probability of measuring a given energy after a local quantum quench. Let us imagine that the system is initially disconnected (i.e. it is in the ground state of $H_A+H_B$), and that at time $t=0$ the interaction between $A$ and $B$ is instantaneously switched on. Then, at time $t>0$, the system evolves with the total Hamiltonian (\ref{eq:totham}).
 If one measures the energy of the system just after the quench, the probability of finding the ground state energy is given by $\big| \langle A\cup B | A\otimes B \rangle \big|^2$, which is the bipartite fidelity. For later convenience, we consider (minus) the logarithm of this quantity
\begin{equation}
	\F_{A,B} = -\ln \left(\big|\langle A\cup B |A\otimes B\rangle\big|^2 \right),
\end{equation}
and call it logarithmic bipartite fidelity (LBF). The symbol $\F_{A,B}$ is chosen because it can be interpreted as a free energy in
 a classical system.

\begin{figure}
\begin{center}
 \includegraphics[width=0.98\textwidth]{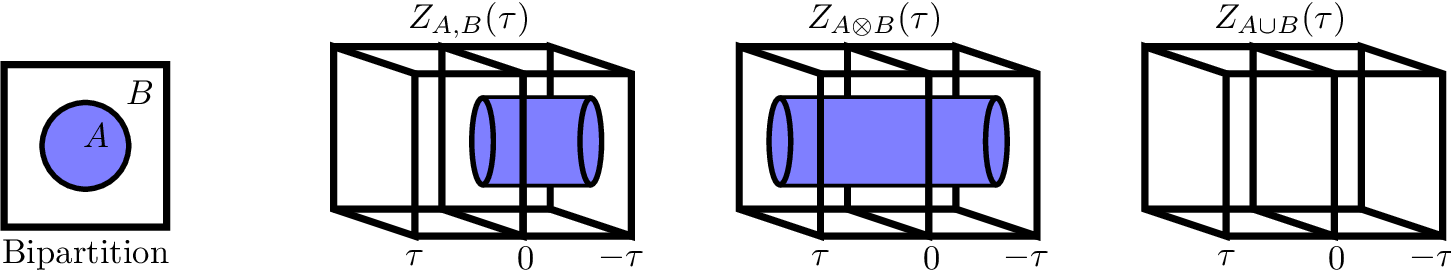}
\caption{Bipartition of a 2D system (on the left), along with the three partition functions of Eq.~\ref{eq:partitionfunctions} 
in $d+1$=$3D$. Region $A$ is in blue when decoupled from $B$.\label{figAB}}
\end{center}
\end{figure}

{\it Free energy and area law}.---
The LBF is nothing but a linear combination of free
 energies of different $d+1$-dimensional systems. Indeed, 
in a euclidean picture, the ground state $|0\rangle$ of a Hamiltonian $H$ can be seen as the result of
 an infinite (imaginary) time evolution starting from any state $|s\rangle$, provided $\left< 0| s\right> \neq 0$
\begin{equation}
 e^{-\tau H}|s\rangle \;
\underset{\tau \to +\infty}{\sim} \;
e^{-\tau E_0} |0\rangle\langle 0 |s\rangle.
\end{equation}
$E_0$ is the ground state energy. Making use of this, our scalar product can be expressed as a ratio of classical $d+1$-dimensional partition functions 
\begin{equation}\label{eq:partitionfunctions}
\langle A\cup B| A\otimes B\rangle=\lim _{\tau \to \infty}\frac{Z_{A,B}(\tau)}{\sqrt{Z_{A\cup B}(\tau)Z_{A\otimes B}(\tau)}}.
\end{equation}
$Z_{A\otimes B} = \left<s\right| e^{-2\tau(H_A+H_B)} \left|s\right> $ is the partition function corresponding to two independent systems $A$ and $B$, $Z_{A\cup B} = \left<s\right| e^{-2\tau(H_A+H_B+H_{A\cup B}^{I})} \left|s\right> $
 is the partition function of the total system, and $Z_{A,B} = \left<s\right| e^{-\tau(H_A+H_B)}e^{-\tau(H_A+H_B+H_{A\cup B}^I)} \left|s\right> $ corresponds to the case when $A$ and $B$ are decoupled from $-\tau$ to $0$,
 and coupled afterwards (see the $d=2$ example in Fig.~\ref{figAB}).
In terms of free energies $f=-\ln Z$, the LBF is then
\begin{equation}
 \mathcal{F}_{A,B}=2f_{A,B}-f_{A \otimes B}-f_{A\cup B}.
\label{eq:freeenergies}
\end{equation}
The different terms in (\ref{eq:freeenergies}) are expected to be extensive in the thermodynamic limit.
 There is a bulk free energy $f_{d+1}$ per unit volume, a ``surface'' free energy $f_d$, and a ``line'' free energy $f_{d-1}$.
 The bulk and surface energies are canceled out by the linear combination, and we get
\begin{equation}
 \F_{A,B}=f_{d-1}L^{d-1}+o(L^{d-1}),
\end{equation}
where $L^{d-1}$ is the ``area'' of the boundary between $A$ and $B$ in the initial $d$-dimensional system.
 This is the area law for the LBF. We expect this to be true for generic systems, as is the case
usually for the EE, however like for the EE \cite{noarealaw}, exceptions are possible.

{\it 1D conformal QCPs}.---
In general, $\mathcal{F}_{A,B}$ should be finite away from a QCP, because
 the correlation length $\xi$ is small and the two ground-states are very
close to each other, except on a thin region of typical size $\xi$.
 At a QCP however,
 this is no longer true and $\mathcal{F}_{A,B}$ can become large.
As the calculation of the overlap boils down to a free energy, the scaling behavior should
be controlled by the conformal symmetry only.
This is indeed the case, and this result constitutes the central point of our work. 
The geometries considered are shown in
Table.~\ref{tab:geometries}, along with the corresponding formulae we derived for the bipartite fidelity.
 For example, geometry (a) consists in a finite chain of length $\ell$ connected to another finite chain of length $L-\ell$.
 When $\ell \ll L$, the scaling of the fidelity takes the following simple form 
\begin{equation}\label{eq:inf}
 \F_{A,B}\sim\frac{c}{8}\ln \ell.
\end{equation}
This result is similar to the one for the EE \cite{cc}: $S\sim (c/6) \ln \ell$. Another simple result is for the symmetric case $\ell=L/2$, where
\begin{equation}\label{eq:symm}
 \F_{A,B}\sim\frac{c}{8} \ln L,
\end{equation}
whereas the EE behaves as $S\sim (c/6)\ln L$ \cite{cc}. There is no general
 relation between the LBF and the EE though. Our analytical results (Table.~\ref{tab:geometries})
 do not match the ones for the EE $S\sim\frac{c}{6}\ln [\frac{L}{\pi}\sin \frac{\pi \ell}{L}]$,
which can be traced back to the fact that the Cardy-Calabrese derivation \cite{cc} of the EE involves a local twist operator,
 whereas the LBF cannot be expressed as a correlator of
a local field. Both quantities have a similar qualitative behavior however.
\begin{center}
\begin{table}
 \begin{tabular}{|m{7cm}|m{7cm}|}
  \hline
\centering Geometry (a)&\centering Geometry (b)\tabularnewline
\hline
\centering
\includegraphics{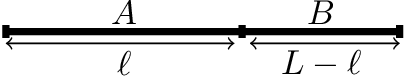}
&
\centering
\vspace{0.1cm}
\includegraphics{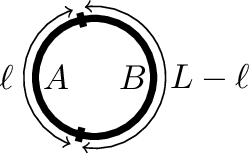}
\tabularnewline
\hline
\vspace{0.2cm}
\centering $\displaystyle{\mathcal{F}_a\sim \frac{c}{8}\Big [ \ln L+g_a(x)+g_a(1-x) \Big]}$&
\vspace{0.2cm}
\centering $\displaystyle{\mathcal{F}_{b}\sim \frac{c}{4}\Big[ \ln L+g_b(x)+g_b(1-x) \Big]}$
\tabularnewline
\vspace{0.4cm}
\centering
$\displaystyle{g_a(x)=\frac{3-3x+2x^2}{3(1-x)}\ln x}$
\vspace{0.2cm}
&
\vspace{0.4cm}
\centering
$\displaystyle{g_b(x)=\frac{3-6x+4x^2}{6(1-x)}\ln x}$
\vspace{0.2cm}
\tabularnewline
\hline
 \end{tabular}
\caption{Geometries considered, along with the leading term for the two LBFs ($\mathcal{F}_a$ and $\mathcal{F}_b$), as a function of $L$ and $x=\ell/L$.
\label{tab:geometries}
}
\end{table}
\end{center}

{\it Conformal field theory derivation}.---
\begin{figure}
\begin{center}
\includegraphics[width=0.85\textwidth]{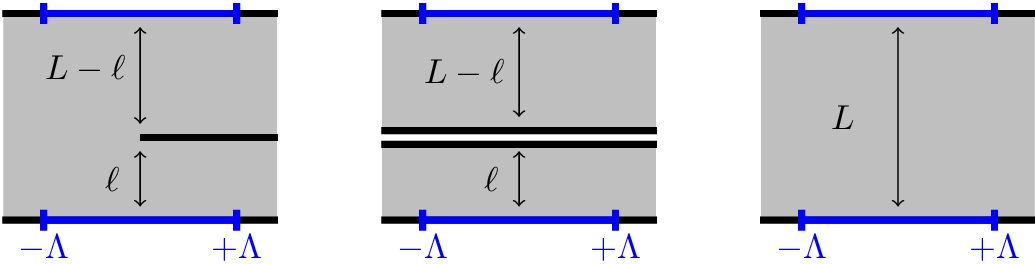}
\end{center}
\caption{The three geometries giving the terms $f_{A,B}$, $f_{A \otimes B}$ and $f_{A \cup B}$ for the case (a)
 in Tab.~\ref{tab:geometries}. The blue intervals $[-\Lambda, + \Lambda]$ are used for the regularization.
\label{fig:3geo}
}
\end{figure}
 The results in
Tab.~\ref{tab:geometries} assume that the boundary conditions
at the boundaries of $A\cup B$ (if any), $A$ and $B$ are conformal
boundary conditions \cite{CFT_intro}. Moreover, for simplicity, we
assume that these boundary conditions are the same everywhere. For
different boundary conditions, the scaling dimensions of the different
boundary condition changing operators \cite{CFT_intro} would modify
our results \cite{DSprep}. With those two assumptions at hand, the two special cases of
geometry (a) given in (\ref{eq:inf})-(\ref{eq:symm}) are straightforward
applications of the celebrated Cardy-Peschel formula \cite{CardyPeschel}: in the
three geometries shown in Fig.~\ref{fig:3geo} there is one corner with angle $2\pi$
and several corners with angle $0$ at infinity. The contributions at infinity cancel,
and we are left with the contribution of the corner with angle $2\pi$, which gives
a total factor of $2 \times \frac{c}{16} \ln \ell$ (or $2 \times \frac{c}{16} \ln L$).

In the general case (a) in Tab.~\ref{tab:geometries}, the calculation goes as follows. $f_{A,B}$,
$f_{A \otimes B}$ and $f_{A \cup B}$ in (\ref{eq:freeenergies}) are
the free energies in the geometries shown in Fig.~\ref{fig:3geo}. Let $w=x+i\, y$ be the complex coordinate such that the lower
 (resp. upper) boundary corresponds to $y=\Im m \, w =0$ (resp. $y= L$). Let us consider the mapping $w \mapsto w + i \,\delta \ell\,$ if $\Im m \, w \in (0, L)$, and $w \mapsto w$ otherwise. Such a mapping keeps $L$ fixed but changes $\ell$ into $\ell + \delta \ell$. The variation of the free energy can be expressed in terms of
the $T_{yy}$ component of the stress-tensor as \cite{CFT_intro}
\begin{equation}
 \delta f_{A,B}  =  \lim_{\Lambda \rightarrow \infty} \frac{\delta \ell}{2 \pi }
\int_{-\Lambda}^{+\Lambda} \Big[\langle T_{yy}^{(y=0)}\rangle - \langle T_{yy}^{(y=L)}\rangle\Big]dx,
\end{equation}
and there are similar expressions for $f_{A \otimes B}$ and $f_{A \cup B}$. Each of these expressions diverges when $\Lambda \rightarrow \infty$,
 but the combination (\ref{eq:freeenergies}) is finite. To evaluate $\delta f_{A,B}$ one
 needs the stress-tensor in the pants-like geometry (Fig.~\ref{fig:3geo} left).
 We use a conformal mapping $z \mapsto w(z)$ from the upper half-plane ${z \in \mathbb{C}, \Im m \, z>0}$ to the pants-like geometry
\begin{equation}
	w(z) \; = \; \frac{\ell}{\pi}\, \ln (1+z) \, + \, \frac{L-\ell}{\pi} \, \ln \left( z \frac{\ell}{L-\ell} -1 \right).
\end{equation}
In the half-plane, one has $\left< T(z)\right>=0$, so using the transformation law of the
stress-tensor \cite{CFT_intro} we get $\left< T(w) \right> = -\frac{c}{12} (dw/dz)^{-2} \{ w,z \}$,
 where $\{ w,z \}=\frac{w'''}{w'}-\frac{3}{2}\left(\frac{w''}{w'}\right)^2$ is the Schwarzian derivative of $w$ with respect to $z$. Since
$\left< T_{yy} \right>  =  \left< T(w) \right> + \left< \bar{T}(\bar{w})\right>$, we find
\begin{equation}
\frac{12\pi}{c} \frac{\delta f_{A,B}}{\delta \ell}  =  
\int_{x_1}^{x_2} dz \,\{w,z\} \left( \frac{dw}{dz}\right)^{-1}-\int_{x_3}^{x_4}dz \,\{w,z\} \left( \frac{dw}{dz}\right)^{-1}
\end{equation}
where $w(x_1)=\Lambda$, $w(x_2)=-\Lambda$, $w(x_3)=\Lambda + i L$, $w(x_4)=-\Lambda+i L$.
In a strip of width $\ell$ the stress-tensor is \cite{CFT_intro} $\left<T(w)\right> = -\frac{ \pi^2 \,c}{24 \, \ell^2}$,
 so $\delta f_{A \otimes B} \,=\, 2 \Lambda \frac{\pi \, c\, \delta \ell}{24} \left(1/(L-\ell)^2-1/\ell^2 \right)$ and $\delta f_{A \cup B}=0$.
 Finally, introducing the aspect ratio $x=\ell/L$ and taking the $\Lambda \rightarrow \infty$ limit in the sum (\ref{eq:freeenergies}),
 we get the following equation for $\mathcal{F}=\mathcal{F}_{A,B}$ :
\begin{eqnarray}\nonumber
 \frac{\delta \mathcal{F}}{\delta x}=\frac{c}{6}\left[
\frac{x^2(2-x)}{2(1-x)^2}\ln x+\frac{x^2-1}{2x}\ln (1-x)+\frac{1}{4}\frac{x-2}{1-x}\right]-\Big(x \to 1-x\Big).
\end{eqnarray}

This can be integrated to give the formula (a) in Tab.~\ref{tab:geometries}. For the periodic case (b), we need a conformal
 transformation which maps the upper half-plane onto a cylinder with two slits
\begin{equation}
 w(z)=\frac{L}{2\pi}\ln \left( 1+\frac{\ell}{L}(z^2-1)\right)-\frac{\ell}{L}\ln z^2,
\end{equation}
and the result follows from a similar calculation.

{\it Numerical checks}.---
We study the XY-chain in transverse field,
 with boundary conditions $\sigma_{L+1}^x=\sigma_{L+1}^y=0$: 
\begin{equation}
 H=-\sum_{i=1}^{L} \left(\frac{1+r}{2}\sigma_i^x \sigma_{i+1}^x+\frac{1-r}{2}\sigma_i^y \sigma_{i+1}^y +h \sigma_i^z\right).
\end{equation}
Two cases are of special interest: $r=0$ and $h=0$ is the critical XX chain, in the universality class of
 the free boson ($c=1$); whereas $r=1$ is the Ising chain in transverse field (ICTF),
 critical at $h=1$ with $c=1/2$.
 Using a Jordan Wigner transformation, $H_A$, $H_B$ and $H_{A\cup B}$ can be recast as free fermions Hamiltonians,
 and diagonalized by a Bogoliubov transformation. Keeping track of the changes of basis,
 the overlap can be expressed as a fermionic correlator,
 and reduced to a $L\times L$ determinant after some algebra.

\begin{figure}[ht]
\begin{center}
 \includegraphics[width=0.33\textwidth,angle=-90]{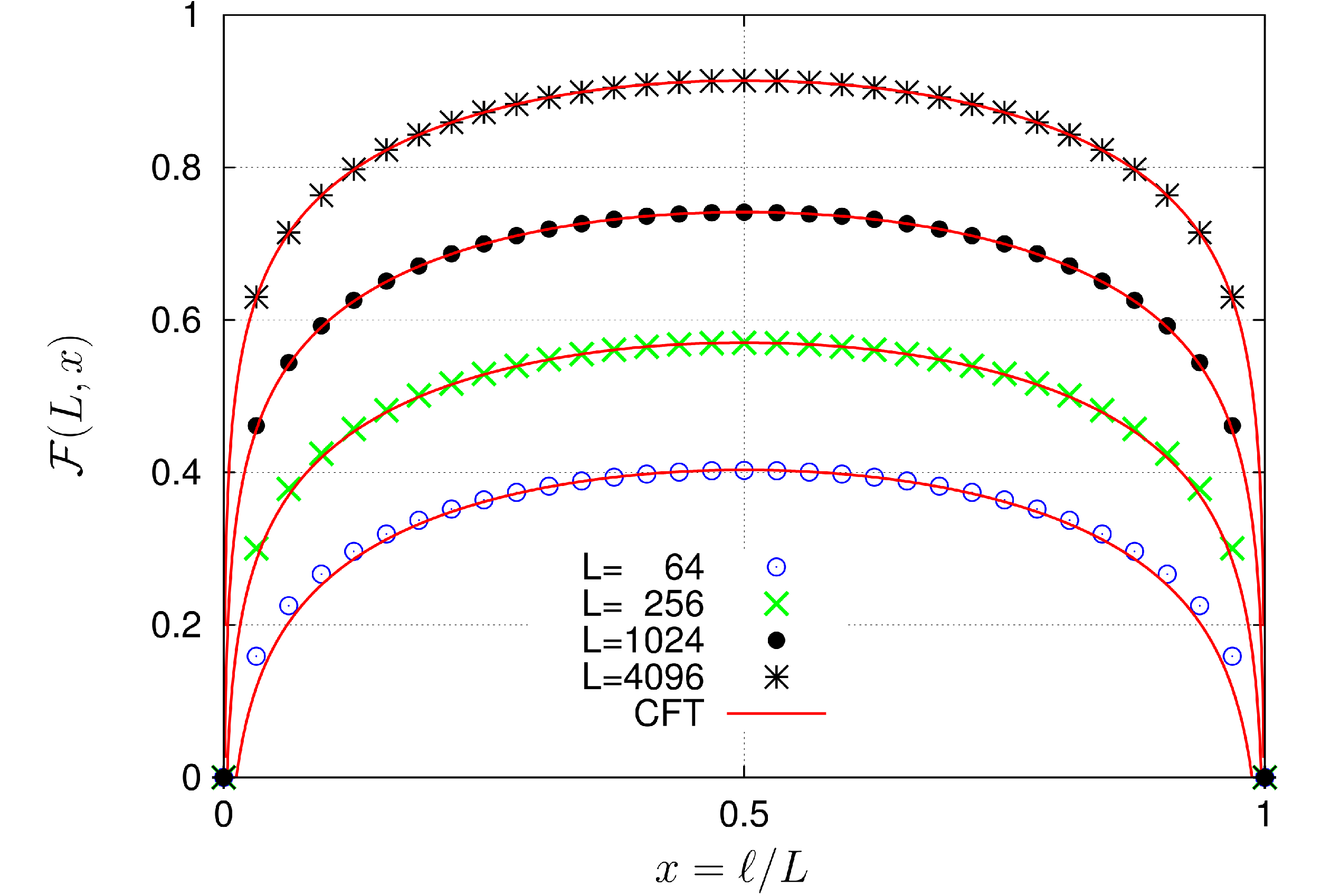}
\includegraphics[width=0.33\textwidth,angle=-90]{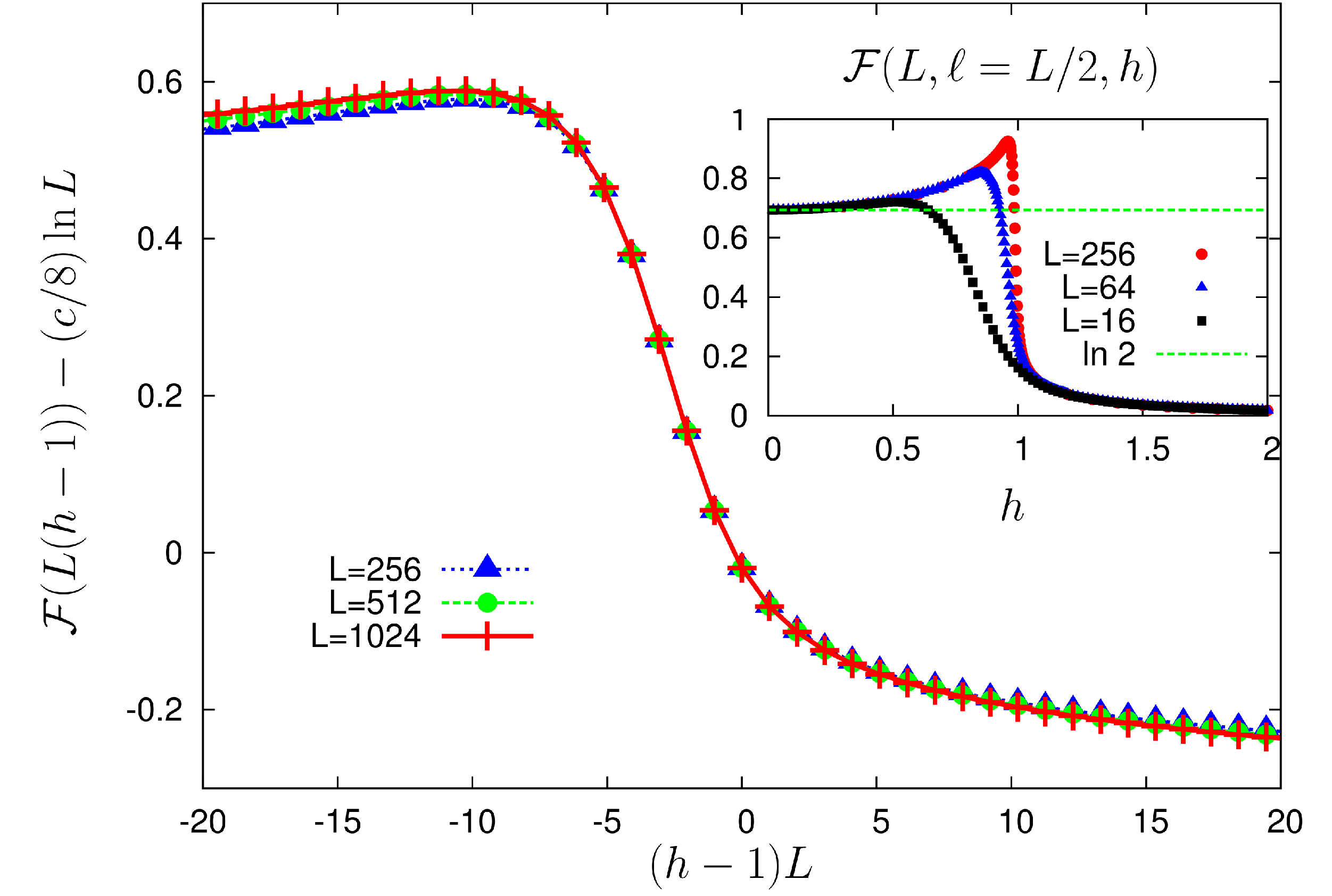}
\end{center}
\caption{Left Panel: XX chain numerical results for the LBF $\mathcal{F}=\mathcal{F}_{A,B}$ in geometry (a).
Right Panel: Geometry (a) with $\ell=L/2$ for the ICTF.
 Rescaled LBF $\F\big(L(h-1)\big)-(c/8)\ln L$ can be seen to collapse onto
 a single universal curve in the vicinity of the critical point. Inset: $\F(L,h)$ as a function of $h$.
\label{fig:checks}
}
\end{figure}

 Results for the XX chain are shown in Fig.~\ref{fig:checks} for geometry (a),
 and agree very well with the CFT prediction. We also checked our formula for geometry (b).

Results for the overlap as a function of $h$ in the ICTF are also shown in Fig.~\ref{fig:checks}. The quantum phase transition
 at $h=1$ can clearly be seen, even with relatively small system sizes. In the vicinity of the critical point,
 the correlation length is known to diverge as $\xi\sim |h-1|^{-1}$, and the rescaled overlaps can be made to
 collapse
onto a universal curve.
Similar to the EE \cite{xx_ising}, we also have the exact relation $\F^{({\rm XX})}(L,\ell)=2\F^{({\rm Ising})}(L/2,\ell/2,h=1)$
 on the lattice.

{\it Time evolution after a local quench}.---
Let us consider again the system in
Tab.~\ref{tab:geometries} (geometry a). It is prepared at time $t=0$ in the
state $\left| A \otimes B \right>$. Then for $t>0$ the two parts $A$ and $B$ interact, and the
system $A\cup B$ evolves with the total Hamiltonian (\ref{eq:totham}). It is well-known that the EE for such
a system grows as \cite{ccquench}  $S \sim \frac{c}{3} \ln t$  for $a \ll v_F t \ll \ell,L~$ 
($a$ is the lattice spacing and $v_F$ the Fermi velocity).
 This logarithmic growth has given rise to speculations about a possible relation between the EE and the statistics
 of fluctuations of the current between the two parts in certain fermionic systems \cite{KlichLevitov},
 which would open the route to an experimental measure of the EE. Here we stress the fact that the time-dependent LBF grows logarithmically as well.
 Actually, the bipartite fidelity in that case is nothing but a Loschmidt echo
\begin{equation}
	\mathcal{L}(t) = \big| \left< A \otimes B \right| e^{i H t} \left| A \otimes B \right> \big|^2
\end{equation}
so the LBF is $\mathcal{F}_{A,B}(t) = - \ln \mathcal{L}(t).$ Its universal behavior can be derived in CFT as follows.
 In imaginary time the scalar product $\left< A \otimes B \right| e^{-\tau H} \left| A \otimes B \right>$ is the partition function
 of a 2D statistical system in a strip with two slits separated by a distance $v_F \tau$. In the limit $v_F \tau \ll \ell,L~$ the
 two slits almost touch each other. Again, the Cardy-Peschel formula shows that
 the contribution to the free energy of each of these corners scales as $\frac{c}{16} \ln v_F \tau$.
 The LBF behaves then as $\mathcal{F}_{A,B}(\tau) \sim \frac{c}{4} \ln \left| \tau \right|$. Going back to real time
$\tau \rightarrow \epsilon -i t$, we find
\begin{equation}
	\mathcal{F}_{A,B}(t) \sim \frac{c}{8} \ln \left( 1 + \frac{t^2}{\epsilon^2} \right) \sim \frac{c}{4} \ln t~.
\end{equation}

{\it 2D conformal QCPs}.---
As discussed before, in $2D$ the bipartite fidelity should scale linearly with the system size $\mathcal{F}=f_1L+o(L)$,
 where $f_1$ depends on the microscopic details of the theory. Universal quantities, if there, have to be looked for in subleading corrections. 
We consider the simple example of a critical quantum dimer wave functions, whose amplitudes are given by the Boltzmann weights of a $2D$ classical
 dimer model. In the continuum limit this wave function is related to a
 free boson CFT with compactification radius $R$~\cite{foo1,alet05}.
 For the geometry of a cylinder of height $L_y\gg L_x$ cut into two parts (see Ref.~\cite{stephan09}),
 the LBF can be expressed using classical partition functions for the dimers~\cite{stephan09}, 
$\mathcal{F}=\ln Z^{DD}(L_x,L_y)-2\ln Z^{DD}(L_x,L_y/2)$.
 $D$ stands for Dirichlet and encode the conformal invariant boundary condition at both
end of the cylinders in the continuum limit.
The first subleading term is a constant related to
the (Dirichlet) Affleck-Ludwig boundary entropy~\cite{AffleckLudwig} $s_D$ computed in Ref.~\cite{fsw}:
\begin{equation}
 \mathcal{F}\sim f_1 L-2s_D=f_1L+\ln R.
\end{equation}
 It would certainly be interesting to study this idea in more complicated models.

 As is the case for the EE \cite{kplw06}, we also speculate that subleading terms in the LBF might be used to identify topological order.
 For certain trial wave-functions, such as Rokhsar-Kivelson triangular lattice quantum dimer and Levin-Wen string net wave-functions,
 the LBF is nothing but the $n \to \infty$ R\'enyi entropy. 
 Then the arguments in \cite{kplw06} would yield the same subleading constant in the LBF and in the EE.
 Another way of looking at this would be to compare the eigenstate of the reduced density matrix associated with its
 largest eigenvalue to the actual ground state of the physical Hamiltonian $H_A$
. For a generic Hamiltonian there is no reason why there should be
any relation between those two states. 
However, for special Hamiltonians associated with trial wave functions, we speculate that they might be closely related to each other.
 We leave this important open question for future studies.

{\it Conclusion}.--- 
We have introduced the LBF of an extended quantum system $A\cup B$, and studied some of its properties.
 We have shown in particular that it generically obeys an area law, and exhibits universal behavior at 1D and 2D QCPs, like the EE.
 We note that its simple definition makes it easier to grasp intuitively than the EE, and 
convenient to study using standard analytical and numerical methods.
 This could be particularly useful in dimension $d>1$, where Quantum Monte Carlo algorithms allow to compute efficiently
 ground state overlaps, as opposed to the von Neumann entropy.
 Therefore, we believe that the LBF can be a useful and general tool in the study of quantum many-body systems.

{\it Acknowledgments}.---
We wish to thank F.~Alet, P.~Calabrese, J.~Cardy, J.-L.~Jacobsen, G.~Misguich, V.~Pasquier and H.~Saleur for
 valuable discussions and encouragements.


\begin{thebibliography}{99}
\bibitem{foo1}
We use the conventions of Ref.~\cite{stephan09} to define $R$.
\end{thebibliography}
\end{document}